\documentclass{article}
\usepackage{ijcai25}

\usepackage{times}
\usepackage{soul}
\usepackage{url}
\usepackage[hidelinks]{hyperref}
\usepackage[utf8]{inputenc}
\usepackage[small]{caption}
\usepackage{graphicx}
\usepackage{amsmath}
\usepackage{amsthm}
\usepackage{booktabs}
\usepackage{algorithm}
\usepackage{algorithmic}
\usepackage[switch]{lineno}
\usepackage{amssymb}
\usepackage{multirow,tabularx}
\usepackage{bm}
\usepackage{color}

\title{Private Transformer Inference in MLaaS: A Survey}

\author{
Yang Li$^{1,2}$
\and
Xinyu Zhou$^{1,2}$\and
Yitong Wang$^{2}$\and
Liangxin Qian$^{2}$\And
Jun Zhao$^2$\\
\affiliations
$^1$Energy Research Institute @ NTU, Interdisciplinary Graduate Programme, Nanyang Technological University, Singapore\\
$^2$College of Computing and Data Science, Nanyang Technological University, Singapore\\
\emails
\{yang048, xinyu003, yitong002, qian0080\}@e.ntu.edu.sg,
junzhao@ntu.edu.sg
}

\begin{document}

\maketitle

\begin{abstract}
Transformer models have revolutionized AI, powering applications like content generation and sentiment analysis. However, their deployment in Machine Learning as a Service (MLaaS) raises significant privacy concerns, primarily due to the centralized processing of sensitive user data. Private Transformer Inference (PTI) offers a solution by utilizing cryptographic techniques such as secure multi-party computation and homomorphic encryption, enabling inference while preserving both user data and model privacy.
This paper reviews recent PTI advancements (2022–2025), highlighting state-of-the-art solutions and challenges. 
We also introduce a structured taxonomy and evaluation framework for PTI, focusing on balancing resource efficiency with privacy and bridging the gap between high-performance inference and data privacy.

\end{abstract}

\section{Introduction}
Transformer models have emerged as game-changers to revolutionize the field of AI. 
For instance, both OpenAI ChatGPT and Microsoft Bing have made the power of transformer-based models widely accessible, democratizing advanced AI capabilities.
These models leverage attention mechanisms~\cite{vaswani2017attention} adeptly to capture long-range dependencies in sequences of input tokens, allowing to model contextual information accurately. Besides, large transformer models are trained on huge quantities of unlabeled textual data and are directly useful for various applications such as sentiment analysis, language translation, content generation, and question answering.

\begin{figure}[t]
    \centering
    \includegraphics[width=0.85\linewidth]{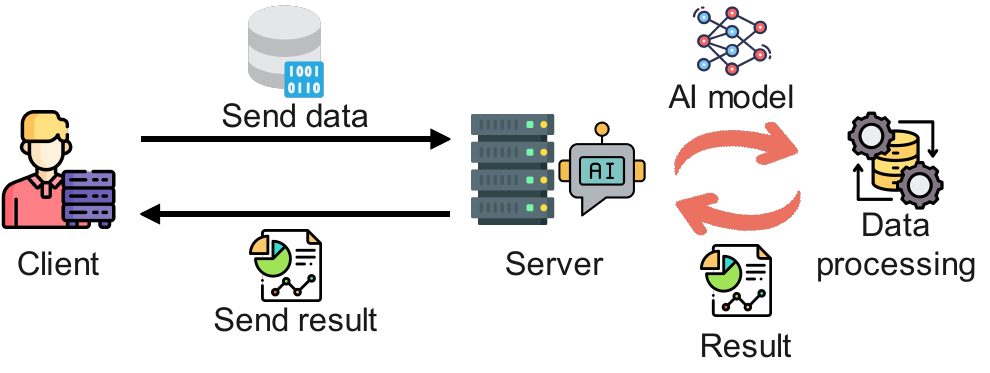}
    \caption{MLaaS without privacy protection}
    \label{fig:MLaaS}
\end{figure}

However, applying large transformers still presents privacy risks~\cite{lund2023chatting,tlili2023if}, particularly with the MLaaS model, where servers provide inference services while users supply data (Figure~\ref{fig:MLaaS}).
For instance, 
OpenAI’s ChatGPT operates through an online platform and APIs, requiring users to transmit private data to the server.
This reliance raises concerns over data misuse, like unauthorized processing, indefinite storage, or resale to third parties. Even with trustworthy servers, centralized data storage remains vulnerable to breaches and insider threats.
Therefore, while MLaaS offers significant convenience and computational power, it also necessitates careful consideration of privacy issues.
Such concerns have prompted actions like Italy’s temporary ban of ChatGPT~\cite{mauran2023whoops,lomas2023italy}. The tension between high-performance transformer services and privacy concerns highlights the need for \textit{Private Transformer Inference} (PTI).
Private inference is a cryptographic protocol that allows for model inference while ensuring that the server gains no knowledge about the users' input, and the users learn nothing about the server's model, apart from inference results. 
Recently, private inference on transformers has been achieved by using private outsourced computation techniques, such as secure Multi-Party Computation (MPC)~\cite{yao1982protocols} and Homomorphic Encryption (HE)~\cite{gentry2009fully}. 
These advancements make PTI highly promising for enabling privacy-preserving AI in practical fields such as banking and healthcare, facilitating secure data analysis while preserving confidentiality. Platforms like Pyte.ai\footnote{\url{https://www.pyte.ai/blog/transformers-based-ai-and-smpc}} and Hugging Face’s Private Hub\footnote{\url{https://huggingface.co/blog/introducing-private-hub}} demonstrate its potential in real-world applications.

To our best knowledge, surveys focusing on PTI do not exist so far, and some recent surveys~\cite{CHITTYVENKATA2023102990,yan2024protecting} only investigate optimizing transformer inference or summarize general privacy issues in Large Language Models (LLMs).
Notably, we are the first to carefully review the findings of state-of-the-art PTI studies and uniquely discuss the improvements, challenges and present future research directions. The rest of the paper is organized as follows: Section~\ref{sec:preliminaries} presents preliminaries; Section~\ref{sec:PTI} provides an overview of PTI; Sections~\ref{sec:linear} and \ref{sec:non-linear} detail solutions to linear and non-linear layers; Section~\ref{sec:experiments} analyzes the experimental results; Section~\ref{sec:conclusion} concludes the paper.

\section{Preliminaries}\label{sec:preliminaries}

\subsection{Transformers} 
Transformer is an encoder-decoder architecture. We focus on the encoder part here, while the decoder can be discussed similarly. 
An encoder consists of a stack of identical blocks, each with a self-attention layer and a feed-forward network.\\
\textbf{Attention layer.} Attention layer first maps the input $\boldsymbol{X}$ into three matrices: the query $\boldsymbol{Q}=\boldsymbol{X}\boldsymbol{W}^Q$, the key $\boldsymbol{K}=\boldsymbol{X}\boldsymbol{W}^K$, the value $\boldsymbol{V}=\boldsymbol{X}\boldsymbol{W}^V$, where $\boldsymbol{W}^Q,\boldsymbol{W}^K,\boldsymbol{W}^V$ are learnable weight matrices. The attention score is computed as follows:
\begin{align}
    \mathrm{Attention}(\boldsymbol{Q},\boldsymbol{K},\boldsymbol{V})= \mathrm{Softmax}\Big(\boldsymbol{Q}\boldsymbol{K}^{T}/\sqrt{d}\Big)\boldsymbol{V}.\label{fun:attention}
\end{align}
Multi-head attention can further extend the above mechanism into $H$ different parallel attention layers.\\
\textbf{Feed-Forward Layer.} A fully connected feed-forward layer consists of two linear transformations with a Gaussian Error Linear Unit (GELU) activation in between:
\begin{align}
    \mathrm{FeedForward}(\boldsymbol{X})= \mathrm{GELU}(\boldsymbol{X}\boldsymbol{W}_1+\boldsymbol{b}_1)\boldsymbol{W}_2+\boldsymbol{b}_2.
\end{align}
In addition to the above blocks, an embedding layer is utilized at the beginning of the model, and each layer is wrapped with residual connections and Layer Normalization (LayerNorm).

\subsection{Cryptographic Primitives}
Various privacy-preserving techniques can be applied to PTI. Here, we focus on two prominent cryptographic primitives: MPC and HE. While MPC and HE may overlap sometimes, we distinguish between them for clarity in this paper. 

\textbf{MPC.} MPC enables multiple parties to jointly compute a function over their inputs while keeping those inputs private. It can be formally described as follows: Consider $n$ parties, $(\mathcal{P}_1,\ldots,\mathcal{P}_n)$, each holding a private input $x_i$. The goal is to jointly compute a function $f(x_1,\ldots,x_n) \rightarrow y$, where $y=(y_1,\ldots,y_n)$. Each party $\mathcal{P}_i$ learns only $y$ (or its portion $y_i$) without gaining any knowledge of others' inputs. 
Protocols such as Secret Sharing (SS) and Oblivious Transfer (OT) are commonly used in MPC, and we brief them as follows:


\begin{itemize}
    \item \textit{Secret Sharing:} SS is a cryptographic technique where a secret $s$ is divided into multiple shares and distributed among $n$ parties. The secret can only be reconstructed when a predefined threshold of shares is combined, and individual shares provide no information about $s$. 
    \item \textit{Oblivious Transfer:} OT allows a sender $\mathcal{S}$ to securely transmit one of $k$ messages $\{m_0,\ldots,m_{k-1}\}$ to a receiver $\mathcal{R}$, who selects the message using a choice bit $b\in[k]$. The receiver $\mathcal{R}$ learns only $m_b$, while the sender $\mathcal{S}$ remains unaware of the receiver's choice.
\end{itemize}
For more details on MPC techniques, readers are suggested to refer to~\cite{evans2018pragmatic,zhao2019secure}.

\textbf{HE.} HE allows computations on encrypted data without decryption, producing results identical to operations on plaintext after decryption. HE schemes use a public key ($\mathrm{pk}$) for encryption and a secret key ($\mathrm{sk}$) for decryption. The public key can be shared freely, while the secret key remains private. Key homomorphic operations are summarized below.
\begin{itemize}
    \item $\mathrm{Enc}(\mathrm{pk},m)\rightarrow c$: On public key $\mathrm{pk}$ and a plaintext $m$, perform encryption to obtain a ciphertext $c$.
    \item $\mathrm{Dec}(\mathrm{sk},c)\rightarrow{m}$: On secret key $\mathrm{sk}$ and a ciphertext $c$, perform decryption to obtain the plaintext message $m$.
    \item \mbox{$\mathrm{Eval}(\mathrm{pk},c_0,\!\ldots\!,c_{k-1},\mathrm{C})\!\rightarrow\!\mathrm{Enc}(\mathrm{pk},\!\mathrm{C}(m_0,\!\ldots\!,m_{k-1}))$:}  
    Evaluate a circuit $\mathrm{C}$ on ciphertexts $c_1,\ldots,c_k$ (i.e., encrypted $m_1,\ldots,m_k$), and output the encrypted result of $\mathrm{C}(m_0,\!\ldots\!,m_{k-1})$.
\end{itemize}
HE schemes could be categorized by the operations used in circuit $\mathrm{C}$ and its computational depth. Due to the page limitation, we do not detail them here.

\section{An Overview of Private Transformer Inference}\label{sec:PTI}

In this section, we first review the threat models, summarize the taxonomy of transformer layers in cryptographic contexts, and finally discuss the challenges of PTI.

\subsection{Threat Models}
The semi-honest (i.e., honest-but-curious) security model is a common assumption in PTI studies.
It assumes that all parties adhere to the established protocols honestly but may attempt to extract additional private information passively. This model is typically employed in scenarios where the parties have a foundational level of trust, ensuring they do not actively disrupt the computation. Under this assumption, participants collaboratively contribute to the computation while maintaining protocol integrity.
Some studies~\cite{akimoto2023privformer,liu2024pptif} involve more than two parties in computation. They assume an honest majority setup, where only a small percentage of participants are semi-honest.

Notably, existing PTI studies are less resistant to malicious attacks~\cite{huang2024secbert}. For instance, if a participant maliciously deviates from the protocol, e.g., by refusing to transmit data, the inference process cannot be completed, but data privacy can still be guaranteed.

\subsection{Taxonomy of Transformer Layers in Cryptographic Contexts}

\begin{figure}[t]
    \centering
\includegraphics[width=\linewidth]{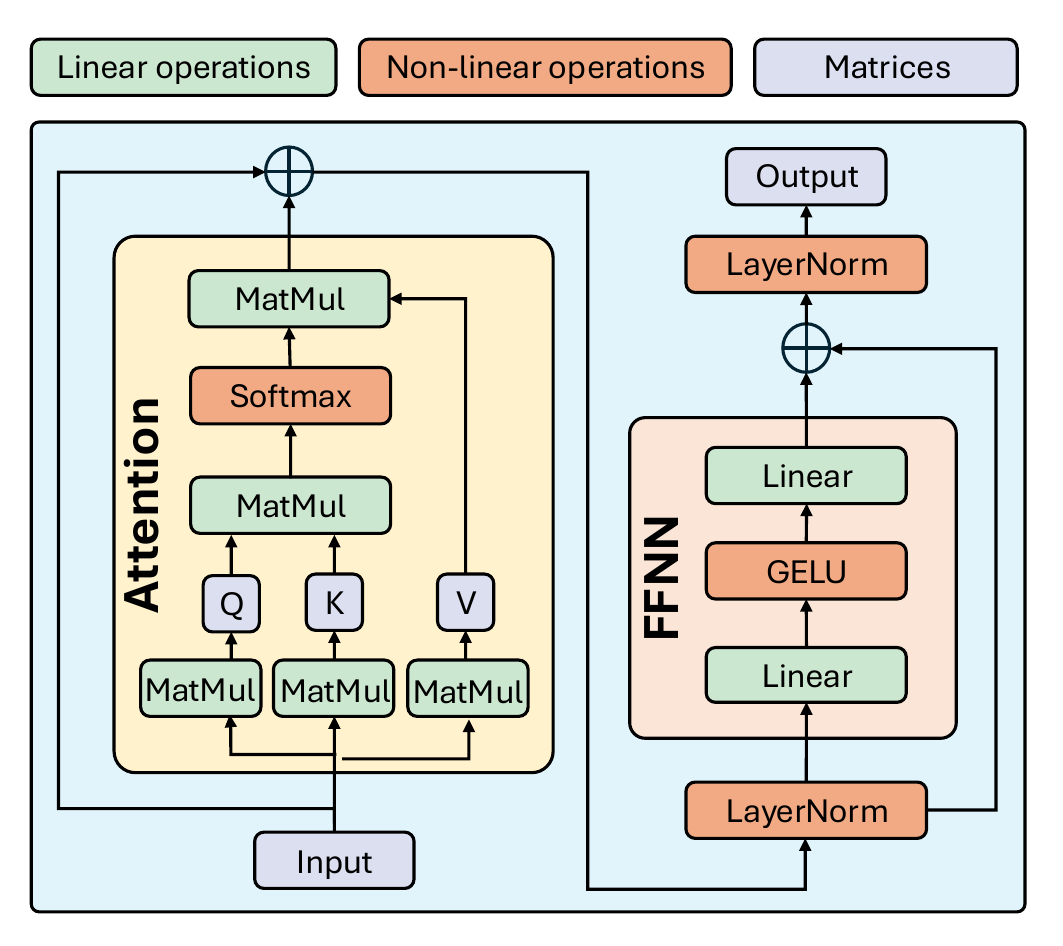}
    \caption{The architecture of a transformer encoder}
    \label{fig:transformer}
\end{figure}

Based on the required cryptographic operations, transformer layers are categorized into linear and non-linear types. 
Figure~\ref{fig:transformer} illustrates the architecture of a basic transformer encoder and highlights these categorizations.

\begin{itemize}
    \item \textit{Linear Layers}: these include embedding, matrix multiplication in attention, and feed-forward layers.
    \item \textit{Non-linear Layers}: these include Softmax, GELU and LayerNorm.
\end{itemize}
We will first summarize the efficient solutions to linear layers in Section~\ref{sec:linear} and then discuss non-linear layers in Section~\ref{sec:non-linear}.

\subsection{Challenges}
This subsection examines the challenges of supporting linear and non-linear layers in the context of cryptography.

Linear layers in PTI focus on matrix multiplication, consisting solely of basic additions and multiplications. Hence, they
are generally compatible with cryptographic techniques. 
However, how to efficiently support large matrix multiplications still remains a significant challenge. 
For example, MPC-based solutions could use techniques such as Beaver triple or OT to support multiplications, but they require substantial data transmission~\cite{lu2023bumblebee}.
In contrast, HE-based solutions are more communication-efficient, as computations can be performed directly on ciphertexts, but they also bring extensive computational overheads due to the complexity of homomorphic operations.

Non-linear layers often involve complex operations beyond basic arithmetic, e.g., $\mathrm{exp}(x)$ in Softmax. Supporting those operations usually requires specific designs and results in greater overhead than basic addition and multiplication.  
Hence, non-linear layers often account for the majority of the overall
overheads in PTI~\cite{hao2022iron,li2022mpcformer,pang2024bolt}. For instance,~\cite{li2022mpcformer} has provided a baseline using MPC where non-linear layers account for more than 85\% of total time (over 50 s). Similarly, \cite{hao2022iron} using HE has also presented that around 60\% of its total execution time (over 280 s) comes from non-linear layers. Thereafter, the significant overhead caused by non-linear layers becomes a key bottleneck for PTI.

\section{Linear Layers}\label{sec:linear}
This section reviews cryptographic protocols for linear layers, focusing on matrix multiplication. The protocols are categorized into two main approaches: MPC-based and HE-based, which will be discussed respectively for clarity.

\subsection{MPC-based MatMul Protocols}

Most MPC implementations for transformers are based on Additive Secret-Sharing (ASS) schemes. In this case, additions can be evaluated locally for ``free'', and multiplications are mainly supported by either Beaver triple~\cite{beaver1992efficient} or Replicated Secret Sharing (RSS)~\cite{araki2016high} method.

Beaver triples are random values $(a,b,c)$ such that $a \cdot b =c$, enabling private multiplications by masking the actual inputs during computation. The triples are also split into shares and distributed to computation parties for privacy.
Studies~\cite{li2022mpcformer,wang2022characterization} introduce an additional trusted third-party dealer to generate triples. \cite{chen2024securetlm} regards the user as a dealer and leaves the computation to multiple servers.
Notably, the triples are independent of inputs and thus are usually generated offline to reduce online overheads. 

RSS is to split a secret into overlapping shares distributed across parties, enabling private multiplication through local computations and limited exchange of intermediate results.
Studies~\cite{dong2023puma,akimoto2023privformer,liu2024pptif} use the RSS scheme for multiplications. Nonetheless, RSS requires at least three parties throughout the online computation process. This means a typical two-party setup (e.g., a server and client) is insufficient, necessitating the involvement of one third-party participant. In such cases, collusion between any two parties compromises security. Hence, the assumption of an honest majority setup is compulsory, which may affect the practicality of applications.

There is no clearly better scheme in comparison between Beaver triples and RSS, as the choice depends on the setup model and application scenarios.
For example, Beaver triples are more efficient during the online phase, whereas RSS performs well on the entire process (offline and online).



\subsection{HE-based MatMul Protocols}

The use of HE for MatMul in Transformers is relatively unexplored. 
While MatMul is inherently compatible with HE, as it involves only additions and multiplications, a straightforward element-wise implementation can be highly inefficient~\cite{chen2022x}. There are currently two mainstream ways to accelerate MatMuls in HE:
\begin{enumerate}
    \item[1)] Encoding plaintexts into SIMD slots.
    \item[2)] Encoding plaintexts into polynomial coefficients.
\end{enumerate}

The SIMD technique supports batching multiple elements into one ciphertext and enables parallel computation within a single operation, significantly reducing the amortized cost. Studies~\cite{pang2024bolt,rovida2024transformer,moon2024thor,zhang2025secure}
have leveraged SIMD in their implementations. 
However, when applied to MatMul, SIMD requires expensive homomorphic rotations to perform the summation. To address this challenge, \cite{pang2024bolt} uses the Baby-Step-Giant-Step (BSGS) to reduce the number of rotations. Similarly, \cite{zhang2025secure} also introduces a slots folding method to reduce the rotation costs.

Studies~\cite{huang2022cheetah,hao2022iron} find that by encoding plaintexts into polynomial coefficients properly, the dot product could directly give the MatMul result and eliminate the need for rotations. Furthermore, \cite{lu2023bumblebee} by Alibaba/Ant proposes a more compact encoding method to alleviate sparsity and save communication. \cite{hou2023ciphergpt} customizes a Vector Oblivious Linear Evaluation (VOLE)-based protocol for MatMuls in GPT, reducing the amortized cost of auto-regressively generating response words.



\section{Non-Linear Layers}\label{sec:non-linear}

Securing non-linear functions presents a significant challenge due to their cryptographic complexity. This section summarizes approaches in existing PTI studies to efficiently realize non-linear layers, including Softmax, GELU and LayerNorm.
\subsection{Softmax}
Softmax is to perform a re-weight normalization of the obtained attention map. 
Given an input vector $\boldsymbol{x}=[x_i\mid_{i\in[d]}]$, the Softmax function on each $x_i$ can be formulated as:
\begin{align}
    \mathrm{Softmax}(x_i)=\frac{\exp{(x_i)}}{\sum_{j=1}^{d}\exp{(x_j)}},\label{equa:softmax}
\end{align}
where the main challenge is efficiently calculating the underlying exponential function $\exp(x)$ and the reciprocal computation $1/x$. The reciprocal computation $1/x$ is relatively well-established and is often treated as a black-box operation~\cite{dong2023puma}. Consequently, recent PTI studies have focused on optimizing the computation of $\exp(x)$.

Several studies employ aggressive crypto-friendly functions to directly replace the $\mathrm{exp}(x)$ in Softmax, which often brings considerable efficiency at the cost of accuracy. Specifically, we present some of their methods as follows:
\begin{align}
    \mathrm{exp}(x) \! \sim \!
    \begin{cases}
        (x+c)^2,&\textnormal{\cite{li2022mpcformer,luo2024secformer}}\\
        \mathrm{RELU}(x), &\textnormal{\cite{zeng2023mpcvit}}\\
        (a x +c)^2,&\textnormal{\cite{zhang2023sal}}\\
        (x +c)^4.&\textnormal{\cite{chen2023rna}}
    \end{cases} \label{equa:appox_softmax1}
\end{align}
It is obvious that substitutions in  (\ref{equa:appox_softmax1}) would make the approximated Softmax differ a lot by numerical values. Hence, the above studies utilize the Knowledge Distillation (KD)~\cite{hinton2006fast} method to bridge the performance gap. However, KD depends on a well-trained teacher model and adds extra computational overhead, which may not be practical.

To maintain the accuracy and eliminate the need for KD, another mainstream solution is to design polynomial approximations for $\mathrm{exp}(x)$. Notably, a key feature of the $\mathrm{exp}(x)$ function is its susceptibility to instability when handling large input values. Hence, most studies will adjust the input range before applying approximations to ensure both stability and computational efficiency.
For instance, \cite{rovida2024transformer} first scales the input into a small range of $[-1,1]$ and then approximates $\mathrm{exp}(x)$ using the Maclaurin series:
\begin{align}
    \mathrm{exp}(x)\approx\sum_{i=0}^6\frac{x^i}{i!}.
\end{align}
To further improve the accuracy and numerical stability, studies~\cite{dong2023puma,lu2023bumblebee,hou2023ciphergpt,zhang2025secure} replace the original input $x_i$ with $x_i-\max{(\boldsymbol{x})}$ to ensure non-positivity, and then design piecewise polynomials with Taylor Series for approximation: 
\begin{align}
    &\exp(x) \approx
    \begin{cases}
        0, &\textnormal{if}~x \leq a\\
        (1+\frac{x}{2^r})^{2^r}, &\textnormal{if}~a\leq x \leq 0
    \end{cases}
\end{align}
where the values of threshold $a$ and polynomial order $r$ differ in the above studies.
Notably, one should be careful about the choice of $r$; too high a polynomial order can bring better accuracy but decrease computational efficiency.


\subsection{GELU}
GELU in Transformers provides smooth and non-linear activation for modeling complex patterns.
The GELU activation for an input $x$ is defined as follows:
\begin{align}
    \mathrm{GeLU}(x)=\frac{x}{2}\big(1+\mathrm{erf}(\frac{x}{\sqrt{2}})\big),
\end{align}
where $\mathrm{erf}(\cdot)$ is the Gaussian error function, expressed as $\mathrm{erf}(x)=\frac{2}{\sqrt{\pi}}\int_0^xe^{-t^2}dt$. Similar to $\mathrm{Softmax}(x)$, solutions to $\mathrm{GeLU}(x)$ in cryptographic context can be mainly categorized into three methods: substitution, approximation, and LUT.

Studies~\cite{chen2022x,park2024powerformer} directly replace GELU with crypto-friendly RELU (i.e., $\max(0,x)$) since support for comparison operations in cryptography is relatively well established~\cite{cheon2020efficient,huang2022cheetah}. MPCFormer~\cite{li2022mpcformer} proposes a more aggressive substitution with a quadratic function:
\begin{align}
    \mathrm{GELU}(x)\sim 0.125x^2+0.25x+0.5.
\end{align}

Other studies~\cite{dong2023puma,lu2023bumblebee,pang2024bolt,zhang2025secure} rely on the polynomials to approximate GELU. Since the GELU function is almost linear with a larger or smaller input, they suggest an efficient low-degree polynomial (e.g.,$n\leq 4$ in \cite{pang2024bolt}) for approximation within the short interval around $0$. A general expression for approximation is shown below:
\begin{align}
    \mathrm{GELU}(x)\approx
    \begin{cases}
        -c,&~\textnormal{if}~x <a\\
        \sum_{i=0}^na_ix^i
           ,
        &~\textnormal{if}~  a \leq x \leq b\\
        x-c,&~\textnormal{if}~b < x\\
    \end{cases}
\end{align}
where $[a,b]$ is a small interval around $0$, $c$ is a small non-negative number (often $0$), and $a_i$ denote the obtained polynomial coefficient from different approximation methods.

\subsection{LayerNorm}

LayerNorm ensures that inputs across different layers have a consistent mean and variance to enhance stability. 
For a given vector $\boldsymbol{x}\in \mathbb{R}^d$, the LayerNorm function is defined as follows:
\begin{align}
    \mathrm{LayerNorm}(x_i) = \frac{(x_i-\mu)}{\sigma}\cdot \gamma+\beta,\label{equa:layernorm}
\end{align}
where $\mu=\sum_{i=1}^dx_i/d$ and $\sigma=\sqrt{\sum_{i=1}^d(x_i-\mu)^2}$ are mean and standard deviation, and $\gamma$ and $\beta$ are affine transform parameters. The main challenge lies in the required reciprocal square root operation of $\sigma$. 

One intuitive solution is to employ well-established protocols for $1/\sqrt{x}$.
Studies \cite{ding2023east,zhang2025secure,park2024powerformer} employ Newton-like methods~\cite{qu2023improvements} to iteratively compute $1/\sqrt{x}$. Similarly,~\cite{luo2024secformer} uses Goldschmidt's method~\cite{markstein2004software} to convert square root inverses into iterations of multiplications. Due to page limitations, we don't detail the protocols here.
In general, the above solutions could maintain the accuracy of LayerNorm but at the cost of considerable overhead.

Some studies manage to avoid non-linear computations by altering the architecture of LayerNorm.
For instance, \cite{chen2022x} directly removes $\mu$ and $\sigma$, leaving the mean and standard deviation achieved by learnable parameters $\gamma$ and $\beta$:
\begin{align}
    \mathrm{LayerNorm}(x_i)\sim x_i \cdot \gamma + \beta.
\end{align}
Similarly, 
\cite{liu2023llms} removes the standard deviation part and then re-trains the model.
In a different way, \cite{rovida2024transformer}  experimentally observes the values of $\mu$ and $\sigma$ to simplify the computation as follows:
\begin{align}
    \mathrm{LayerNorm}(x_i)\approx(x_i-E_p) \cdot (V_p\gamma) + \beta,
\end{align}
where $E_p$ and $V_p$ are precomputed values for $\mu$ and $1/\sigma$. Notably, these methods simplify computations at the expense of LayerNorm accuracy, and therefore often require re-training to recover model accuracy.


\section{Reported Experiments}\label{sec:experiments}
This section compares the reported experimental results presented in selected studies. A summary of their evaluations is presented in Table~\ref{tab:performance}. 

\begin{table*}[ht!]
\centering
\caption{Resource requirements, tasks performed, dataset performance, and runtime.}
\resizebox{\textwidth}{!}{
\begin{tabular}{c|c|c|c|c|c|ccc|c}
\toprule
\multirow{2}{*}{Study} & \multirow{2}{*}{Techniques} & \multirow{2}{*}{Model}     & \multirow{2}{*}{Dataset}  & \multirow{2}{*}{Comm.} & \multirow{2}{*}{Comm. Set.} & \multicolumn{3}{c|}{Performance}  & \multirow{2}{*}{Runtime}  \\  \cline{7-9}
&   &   & &   &  & \multicolumn{1}{c|}{Plain} & \multicolumn{1}{c|}{Enc.} & Loss~$\downarrow$ &    \\ \hline

\multirow{4}{*}{\cite{li2022mpcformer}}  & \multirow{4}{*}{MPC} & \multirow{4}{*}{BERT-Base} 
& QNLI   & \multirow{4}{*}{12.089 GB}   & \multirow{4}{*}{(5 Gbps, 1 ms)}
&\multicolumn{1}{c|}{91.7 \%}    & \multicolumn{1}{c|}{90.6 \%}   & 1.1 \%   & \multirow{4}{*}{55.320 s}  \\
&   & & MRPC   &  &     & \multicolumn{1}{c|}{90.3 \%}     & \multicolumn{1}{c|}{88.7 \%}    & 1.6 \%    &  \\ 
&   & & RTE   &  &     & \multicolumn{1}{c|}{69.7 \%}     & \multicolumn{1}{c|}{64.9 \%}    & 4.8 \%    & \\
& &  & STS-B   &  &     & \multicolumn{1}{c|}{89.1 \%}     & \multicolumn{1}{c|}{80.3 \%}    & 8.8 \%    &       \\\hline

\multirow{9}{*}{\cite{dong2023puma}}  & \multirow{9}{*}{MPC} & \multirow{3}{*}{BERT-Base} & CoLA  & \multirow{3}{*}{10.773 GB} &\multirow{9}{*}{(5 Gbps, 1 ms)}   & \multicolumn{1}{c|}{61.6 \%}    & \multicolumn{1}{c|}{61.3 \%}   & 0.3 \%   & \multirow{3}{*}{33.913 s}         \\
&  & & RTE   &    &    & \multicolumn{1}{c|}{70.0 \%}     & \multicolumn{1}{c|}{70.0 \%}    & 0.0 \%    & \\ 
&  & & QNLI   &    &    & \multicolumn{1}{c|}{91.6 \%}     & \multicolumn{1}{c|}{91.6 \%}    & 0.0 \%    &   \\\cline{3-4}\cline{5-5}\cline{7-10}
& & Roberta-Base  & -      & 11.463 GB &   & \multicolumn{1}{c|}{- }     & \multicolumn{1}{c|}{-}   & -   & 41.641 s  \\\cline{3-4}\cline{5-5}\cline{7-10}
& & Bert-Large  & - &    27.246 GB &   & \multicolumn{1}{c|}{- }     & \multicolumn{1}{c|}{-}   & -   & 73.720 s  \\\cline{3-5}\cline{7-10}
 & & GPT2-Base  & \multirow{3}{*}{Wiki.-103}   & 3.774 GB &  & \multicolumn{1}{c|}{16.284 }     & \multicolumn{1}{c|}{16.284}   & 0.000   & 15.506 s  
\\\cline{3-3}\cline{5-5}\cline{7-10}
&  &GPT2-Medium  &   &     7.059 GB &   & \multicolumn{1}{c|}{12.536 }     & \multicolumn{1}{c|}{12.540}   & -0.004   & 30.272 s  \\\cline{3-3}\cline{5-5}\cline{7-10}
&  & GPT2-Large  &  &     11.952 GB  &  & \multicolumn{1}{c|}{10.142}     & \multicolumn{1}{c|}{10.161}   & -0.019   & 54.154 s  \\\cline{3-5}\cline{7-10}
& & LLaMA-7B  & -   & 1.794 GB   &     & \multicolumn{1}{c|}{-}  &\multicolumn{1}{c|}{-} & \multicolumn{1}{c|}{-}   & 200.473 s  \\
\hline

\multirow{13}{*}{\cite{gupta2023sigma}}  & \multirow{13}{*}{MPC} & \multirow{3}{*}{BERT-Tiny} & SST2 & \multirow{3}{*}{0.02 GB} &\multirow{13}{*}{(9.4 Gbps, 0.05 ms)}   & \multicolumn{1}{c|}{81.19 \%}    & \multicolumn{1}{c|}{81.42 \%}   & -0.23 \%   & \multirow{3}{*}{0.09 s}         \\
&  & & MRPC   &    &    & \multicolumn{1}{c|}{72.54 \%}     & \multicolumn{1}{c|}{72.79 \%}    & -0.25 \%    & \\ 
&  & & QNLI   &    &    & \multicolumn{1}{c|}{81.64 \%}     & \multicolumn{1}{c|}{81.73 \%}    & -0.09 \%    &   \\\cline{3-4}\cline{5-5}\cline{7-10}
& & \multirow{3}{*}{BERT-Base}  & SST2      & \multirow{3}{*}{0.99 GB} &   & \multicolumn{1}{c|}{90.59 \%}     & \multicolumn{1}{c|}{90.25 \%}    & 0.34 \%   & \multirow{3}{*}{1.84 s}  \\
&  & & MRPC   &    &    & \multicolumn{1}{c|}{84.31 \%}     & \multicolumn{1}{c|}{83.82 \%}    & 0.49 \%    & \\
&  & & QNLI   &    &    & \multicolumn{1}{c|}{88.72 \%}     & \multicolumn{1}{c|}{89.03 \%}    & -0.31 \%    &
\\\cline{3-4}\cline{5-5}\cline{7-10}
& & \multirow{3}{*}{BERT-Large}  & SST2      & \multirow{3}{*}{2.63 GB} &   & \multicolumn{1}{c|}{88.99 \%}     & \multicolumn{1}{c|}{88.99 \%}    & 0.0 \%   & \multirow{3}{*}{4.73 s}  \\
&  & & MRPC   &    &    & \multicolumn{1}{c|}{78.67 \%}     & \multicolumn{1}{c|}{78.92 \%}    & -0.25 \%    & \\
&  & & QNLI   &    &    & \multicolumn{1}{c|}{92.23 \%}     & \multicolumn{1}{c|}{92.31 \%}    & -0.08 \%    &
\\\cline{3-4}\cline{5-5}\cline{7-10}
& & GPT-2  & Lambada &  0.82 GB &   & \multicolumn{1}{c|}{32.46 \%}     & \multicolumn{1}{c|}{33.28 \%}    & -0.82 \%   & 1.61 s  \\\cline{3-5}\cline{7-10}
 & & GPT-Neo  & Lambada   & 4.02 GB &  & \multicolumn{1}{c|}{57.46 \%}     & \multicolumn{1}{c|}{57.81 \%}    & -0.35 \%  & 7.43 s  
\\\cline{3-4}\cline{5-5}\cline{7-10}
&  &LLaMA2-7B  & Lambada  &     12.35 GB &   & \multicolumn{1}{c|}{70.17 \%}     & \multicolumn{1}{c|}{70.01 \%}    & 0.16 \%    & 27.01 s  \\\cline{3-4}\cline{5-5}\cline{7-10}
& & LLaMA2-13B  & Lambada   & 19.33 GB  & & \multicolumn{1}{c|}{73.14 \%}     & \multicolumn{1}{c|}{72.98 \%}    & 0.16 \%   & 44.13 s  \\
\hline

\multirow{6}{*}{\cite{zheng2023primer}}  & \multirow{6}{*}{MPC} & \multirow{2}{*}{BERT-Tiny} & MRPC & \multirow{2}{*}{0.9 GB} &\multirow{6}{*}{(100 Mbps, 2.3 ms)}   & \multicolumn{1}{c|}{-}    & \multicolumn{1}{c|}{79.3 \%}   & -   & \multirow{2}{*}{10.6 s}         \\
&  & & SST-2   &    &    & \multicolumn{1}{c|}{-}     & \multicolumn{1}{c|}{88.2 \%}  & -   &   \\\cline{3-4}\cline{5-5}\cline{7-10}
& & \multirow{2}{*}{BERT-Base}  & SST2      & \multirow{2}{*}{3.6 GB} &   & \multicolumn{1}{c|}{-}     & \multicolumn{1}{c|}{86.3 \%}    & -   & \multirow{2}{*}{35.4 s}  \\
&  & & QNLI   &    &    & \multicolumn{1}{c|}{-}     & \multicolumn{1}{c|}{92.5 \%}    & -    &
\\\cline{3-4}\cline{5-5}\cline{7-10}
& & \multirow{2}{*}{BERT-Large}  & SST2      & \multirow{2}{*}{7.9 GB} &   & \multicolumn{1}{c|}{-}     & \multicolumn{1}{c|}{87.6 \%}    & -   & \multirow{2}{*}{91.6 s}  \\
&  & & QNLI   &    &    & \multicolumn{1}{c|}{-}     & \multicolumn{1}{c|}{93.5 \%}    & -   &
\\\hline

\multirow{3}{*}{\cite{zeng2023mpcvit}}  & \multirow{3}{*}{MPC} & \multirow{3}{*}{CCT} & CIFAR-10 & \multirow{3}{*}{-}  & \multirow{3}{*}{(44 Mbps, 40 ms)} & \multicolumn{1}{c|}{95.56 \%} & \multicolumn{1}{c|}{94.27 \%}   & 1.29 \%   & 50.94 s \\

&         &           & CIFAR-100   &   &    & \multicolumn{1}{c|}{77.36 \%}     & \multicolumn{1}{c|}{77.76 \%}    & -0.40 \%    &  51.33 s \\ 
&         &           & Tiny-ImageNet   &   &    & \multicolumn{1}{c|}{61.60 \%}     & \multicolumn{1}{c|}{63.45 \%}    & -2.35\%    & 74.34 s  \\\hline

\multirow{3}{*}{\cite{zhang2023sal}}  & \multirow{3}{*}{MPC} & \multirow{3}{*}{CCT} & CIFAR-10 & \multirow{3}{*}{-}  & \multirow{3}{*}{-} & \multicolumn{1}{c|}{95.56 \%} & \multicolumn{1}{c|}{95.92 \%}   & -0.36 \%   & 40.65 s \\

&         &           & CIFAR-100   &   &    & \multicolumn{1}{c|}{77.36 \%}     & \multicolumn{1}{c|}{77.86 \%}    & -0.50 \%    &  38.97 s \\ 
&         &           & Tiny-ImageNet   &   &    & \multicolumn{1}{c|}{61.60 \%}     & \multicolumn{1}{c|}{63.49 \%}    & -2.39\%    & 66.21 s  \\\hline

\multirow{3}{*}{\cite{chen2023rna}}  & \multirow{3}{*}{MPC} & \multirow{3}{*}{CCT} & CIFAR-10 & \multirow{3}{*}{-}  & \multirow{3}{*}{-} & \multicolumn{1}{c|}{95.56 \%} & \multicolumn{1}{c|}{94.97 \%}   & 0.59 \%   & 14.04 s \\

&         &           & CIFAR-100   &   &    & \multicolumn{1}{c|}{77.36 \%}     & \multicolumn{1}{c|}{79.04 \%}    & -1.68 \%    &  14.13 s \\ 
&         &           & Tiny-ImageNet   &   &    & \multicolumn{1}{c|}{61.60 \%}     & \multicolumn{1}{c|}{65.86 \%}    & -4.26\%    & 44.27 s  \\\hline

\multirow{4}{*}{\cite{luo2024secformer}} &
\multirow{4}{*}{MPC} &
\multirow{2}{*}{BERT-Base}  & MRPC   & \multirow{2}{*}{23.593 GB} &\multirow{4}{*}{(10 Gbps)}   &  \multicolumn{1}{c|}{90.3 \%}     & \multicolumn{1}{c|}{89.2 \%}    & 1.1 \%   & \multirow{2}{*}{19.513 s}          \\
&         &           & STS-B   &   &    & \multicolumn{1}{c|}{89.1 \%}     & \multicolumn{1}{c|}{87.4 \%}    & 1.7 \%    &  \\\cline{3-5}\cline{7-10}
& & \multirow{2}{*}{BERT-Large}    & MRPC  & \multirow{2}{*}{50.364 GB}      &  & \multicolumn{1}{c|}{90.6 \%}     & \multicolumn{1}{c|}{88.7 \%}    & 1.9 \%    & \multirow{2}{*}{39.089 s}                          \\
&         &           & STS-B   &   &    & \multicolumn{1}{c|}{90.2 \%}     & \multicolumn{1}{c|}{89.2 \%}    & 1.0 \%    &  \\
\hline

\multirow{3}{*}{\cite{hao2022iron}} & \multirow{3}{*}{MPC+HE} & \multirow{3}{*}{BERT-Base} & SST-2   & \multirow{3}{*}{280.99 GB}  & \multirow{3}{*}{(3 Gbps, 0.8 ms)}  & \multicolumn{1}{c|}{92.36 \%}    & \multicolumn{1}{c|}{92.77 \%}   & -0.41 \%   & \multirow{3}{*}{475 s}         \\
&  & & MRPC   &    &     & \multicolumn{1}{c|}{90.00 \%}     & \multicolumn{1}{c|}{89.87 \%}    & 0.13 \%    & \\ 
& &   & STS-B   &    &     & \multicolumn{1}{c|}{89.62 \%}     & \multicolumn{1}{c|}{89.41 \%}    & 0.21 \%    &  \\\hline

\multirow{7}{*}{\cite{lu2023bumblebee}} & \multirow{7}{*}{MPC+HE} & \multirow{3}{*}{BERT-Base} & QNLI   & \multirow{3}{*}{-} & \multirow{7}{*}{(1 Gbps, 0.5 ms)}   & \multicolumn{1}{c|}{90.30 \%}    & \multicolumn{1}{c|}{90.20 \%}   & 0.10 \%   & \multirow{3}{*}{-}         \\
&          &          & RTE   &    &    & \multicolumn{1}{c|}{70.04 \%}     & \multicolumn{1}{c|}{70.04 \%}    & 0.00 \%    &  \\
&          &          & CoLA   &      &  & \multicolumn{1}{c|}{61.57 \%}     & \multicolumn{1}{c|}{60.82 \%}    & 0.75 \%    &  \\\cline{3-4}\cline{5-5}\cline{7-10}
& & BERT-Large  & -  &   20.85 GB &  & \multicolumn{1}{c|}{-}     & \multicolumn{1}{c|}{-}   & -  & 404.4 s  \\\cline{3-4}\cline{5-5}\cline{7-10}
& & LLaMA-7B  & -      & 6.82 GB &  & \multicolumn{1}{c|}{-}     & \multicolumn{1}{c|}{-}   & -   & 832.2 s  \\\cline{3-5}\cline{7-10}
& & GPT2-Base  & -    & 1.94 GB &  & \multicolumn{1}{c|}{-}     & \multicolumn{1}{c|}{-}   & -  & 55.2 s  \\\cline{3-5}\cline{7-10}
& & ViT-Base  & ImageNet    & 14.44 GB &  & \multicolumn{1}{c|}{89.44 \%}     & \multicolumn{1}{c|}{89.13 \%}   & 0.31 \%   & 234 s  \\
\hline

\multirow{4}{*}
{\cite{pang2024bolt}} & \multirow{4}{*}{MPC+HE} & \multirow{4}{*}{BERT-Base}   & SST-2   & \multirow{4}{*}{25.74 GB} &\multirow{4}{*}{(3 Gbps, 0.8 ms)}  & \multicolumn{1}{c|}{92.36 \%}    & \multicolumn{1}{c|}{92.78 \%}   & -0.42 \%   & \multirow{4}{*}{185 s}          \\
&          &         & MRPC   &  &       & \multicolumn{1}{c|}{90.00 \%}     & \multicolumn{1}{c|}{89.95 \%}    & 0.05 \%    &  \\ 
&          &           & RTE   &  &      & \multicolumn{1}{c|}{69.70 \%}     & \multicolumn{1}{c|}{69.31 \%}    & 0.39 \%    &  \\
&          &           & STS-B  &  &      & \multicolumn{1}{c|}{89.62 \%}     & \multicolumn{1}{c|}{88.44 \%}    & 1.18 \%    &  \\ \hline

\multirow{2}{*}{\cite{chen2022x}}   & \multirow{2}{*}{HE} & \multirow{2}{*}{BERT-Tiny} & SST-2 & \multirow{2}{*}{-}   & \multirow{2}{*}{-} & \multicolumn{1}{c|}{82.45 \%}    & \multicolumn{1}{c|}{82.11 \%}   & 0.34 \%   & \multirow{2}{*}{$\approx$ 4700 s}         \\
&  &  & STS-B   &   &   & \multicolumn{1}{c|}{72.83 \%}     & \multicolumn{1}{c|}{68.39 \%}    & 4.44 \%    &   \\\hline

\cite{rovida2024transformer}   & HE & BERT-Tiny & SST-2 & -  & - & \multicolumn{1}{c|}{83.7 \%} & \multicolumn{1}{c|}{79.0 \%}   & 4.7 \%   & 214 s \\\hline

\cite{moon2024thor}   & HE & BERT-Base & MRPC & -  & - & \multicolumn{1}{c|}{85.29 \%} & \multicolumn{1}{c|}{84.80 \%}   & 0.49 \%   & 625.8 s \\\hline

\multirow{3}{*}{\cite{zimerman2024power}}  & \multirow{3}{*}{HE} & \multirow{3}{*}{Roberta-Base} & SST-2 & \multirow{3}{*}{-}  & \multirow{3}{*}{-} & \multicolumn{1}{c|}{94.80 \%} & \multicolumn{1}{c|}{93.35 \%}   & 1.45 \%   & \multirow{3}{*}{$\approx$ 400 s} \\

&         &           & QNLI   &   &    & \multicolumn{1}{c|}{92.80 \%}     & \multicolumn{1}{c|}{91.62 \%}    & 1.18 \%    &   \\ 
&         &           & MNLI   &   &    & \multicolumn{1}{c|}{87.60 \%}     & \multicolumn{1}{c|}{86.93 \%}    & 0.67 \%    &  \\\hline

\multirow{6}{*}{\cite{zhang2025secure}} &
\multirow{6}{*}{HE} &
\multirow{3}{*}{BERT-Base}  & RTE   & \multirow{6}{*}{0.16 GB} &\multirow{6}{*}{(100 Mbps, 80 ms)}   & \multicolumn{1}{c|}{70.04 \%}    & \multicolumn{1}{c|}{69.88 \%}   & 0.16 \%   & \multirow{3}{*}{857 s}          \\
&         &           & SST-2   &   &    & \multicolumn{1}{c|}{92.36 \%}     & \multicolumn{1}{c|}{92.11 \%}    & 0.25 \%    &   \\ 
&         &           & QNLI   &   &    & \multicolumn{1}{c|}{90.30 \%}     & \multicolumn{1}{c|}{89.92 \%}    & 0.38 \%    &  \\\cline{3-4}\cline{7-10}
& & \multirow{3}{*}{LLaMA-3B}    & RTE  &      &  & \multicolumn{1}{c|}{82.75 \%}     & \multicolumn{1}{c|}{81.24 \%}    & 1.51 \%    &   \multirow{3}{*}{1088 s}                       \\
&         &           & SST-2   &   &    & \multicolumn{1}{c|}{94.94 \%}     & \multicolumn{1}{c|}{94.46 \%}    & 0.48 \%    &   \\ 
&         &           & QNLI   &   &    & \multicolumn{1}{c|}{90.70 \%}     & \multicolumn{1}{c|}{90.20 \%}    & 0.50 \%    &   \\ 

\bottomrule
\end{tabular}
}
\label{tab:performance}
\end{table*}

\subsection{Models and Datasets} 
Popular transformers for PTI studies mainly include the BERT family (e.g., Bert-Tiny, Bert-Base, Roberta-Base) and the GPT family (e.g., GPT2-Base). Some studies have also attempted implementing other transformer models, such as LLaMA-7B~\cite{touvron2023llama}, ViT-Base~\cite{dosovitskiy2020image}, and its variant CCT~\cite{hassani2021escaping}. Transformers with more complex structures and a larger number of parameters tend to take longer to do inference, but they often perform better~\cite{jiao2019tinybert}.

Most PTI studies evaluate performance using the GLUE benchmark~\cite{wang2018glue}, a standard for BERT and GPT-based transformers, which requires models to process single-sentence and sentence-pair inputs for predictions.
It contains three NLU tasks and nine corresponding corpora: single-sentence tasks (CoLA, SST$-$2), similarity and paraphrase tasks (MRPC, STS$-$B, QQP), and inference tasks (MNLI, QNLI, RTE, WNLI).  
Studies \cite{dong2023puma,zhang2025secure} evaluate GPT2 on Wikitext-103 V1~\cite{merity2016pointer} and CBT-CN. ViT-Base is designed for image processing, and thus BumbleBee~\cite{lu2023bumblebee} uses the ImageNet dataset for its evaluation. A series of studies~\cite{zeng2023mpcvit,zhang2023sal,chen2023rna} focusing on CCT are evaluated on CIFAR and Tiny-ImageNet.

\subsection{Communication Overhead}
According to Table~\ref{tab:performance}, it is obvious that the studies employing MPC usually require considerable communication overhead.
This is because their cornerstone MPC techniques (e.g., SS and OT) all require multiple rounds of communication between parties for multiplication and other nonlinear operations, thus causing significant communication costs. Early study~\cite{hao2022iron} using ASS requires $280.99$ GB for evaluating BERT-Base, while study~\cite{gupta2023sigma} requires only $0.99$ GB on the same model by using a more efficient Function Secret Sharing (FSS) scheme.

In contrast, studies only using HE techniques~\cite{rovida2024transformer,zimerman2024power,zhang2025secure} often require minimum communication. Once the encrypted input from the client is transmitted to the server, the remaining computation will be completed by the server only, which is a non-interactive process. For instance, NEXUS~\cite{zhang2025secure} requires only $0.16$ GB for evaluating the BERT-Base model, which is lower than any other MPC studies. Other studies only using HE do not report the communication overhead, as the main bottleneck in them comes from computation. We will detail this in the next subsection.


\subsection{Runtime}
To ensure fairness, we report the runtime performance of selected studies on CPU only in Table~\ref{tab:performance}, even though some of them also support GPU implementation. However, as most of these studies are not open source, differences in experimental platforms may still affect the runtime performance. Due to page limitations, we only provide their communication setups for readers' reference, and more detailed computation setups can be found in their manuscripts.

Overall, PTI solutions relying only on MPC tend to achieve faster runtimes than HE-only and hybrid (MPC+HE) ones. Those solutions mainly use secret-sharing schemes with low computation complexity but high communication overheads. Consequently, they often deliver the fastest runtimes, particularly in communication-rich environments such as LAN settings. Under a setup with a bandwidth of $9.4$ Gbps and a latency of $0.05$ ms, \cite{gupta2023sigma} demonstrated that BERT-Base could be evaluated in just $1.84$ s.
Hybrid solutions use HE for linear computations to reduce communication overhead, but they also increase runtime due to the computational complexity of homomorphic operations. For instance, \cite{pang2024bolt} evaluates BERT-Base in \mbox{$185$ s}.
HE-only solutions require the longest runtime. 
Latest studies~\cite{zimerman2024power,moon2024thor,zhang2025secure} all require over $400$ s for evaluation on the BERT-Base model. Even for the smaller BERT-Tiny model, \cite{rovida2024transformer} still requires over $200$ s. In particular, the most consuming part of those studies is the \textit{Bootstrapping} operation, which is often used to ``refresh'' a ciphertext to reduce noise. The bootstrapping part in \cite{zhang2025secure} and \cite{moon2024thor} accounts for $37.72\%$ and $53.96\%$ of the total runtime, respectively.

\subsection{Accuracy}

This subsection reviews the accuracy performance of different PTI solutions.
Specifically, linear computations (e.g., matrix multiplication) are often well supported by cryptographic techniques. Hence, most of the accuracy drop reported in Table~\ref{tab:performance} comes from the treatment of non-linear layers. For example, an early study~\cite{li2022mpcformer} employed aggressive substitutions for the Softmax and GELU functions, leading to a substantial accuracy drop of $8.8\%$ on the STS-B dataset. In contrast, a more recent study~\cite{pang2024bolt} utilized refined polynomial approximations, reducing the accuracy loss to just $1.18\%$ on the same dataset. Notably, knowledge distillation techniques could effectively cure the accuracy drop due to approximation operations. Studies~\cite{zeng2023mpcvit,zhang2023sal,chen2023rna} using knowledge distillation even achieved better performances than the plaintext model on CIFAR-100 and Tiny-ImageNet datasets.

\section{Conclusion and Future Directions}\label{sec:conclusion}
In this paper, we have comprehensively reviewed the existing cryptographic solutions for private transformer inference. To the best of our knowledge, this is the first review to categorize these solutions within a structured taxonomy, providing an in-depth discussion. In addition, we also offer a detailed performance comparison of various methods from different perspectives. Despite these advancements, several open challenges remain unresolved, which we summarize as follows:

\textbf{GPU Acceleration.}
One important limitation of current PTI solutions is the lack of GPU acceleration support. Cryptographic operations, particularly those in HE, are computationally expensive. Hence, leveraging GPUs for those operations could dramatically improve the private inference speed~\cite{zhang2025secure}. 
Most PTI studies rely on popular CPU-based cryptographic libraries (e.g., EzPC and SEAL) for implementations. Only a few studies~\cite{gupta2023sigma,zhang2025secure} have explored GPU-accelerated solutions, achieving significant speedups of $15.0\times$ and $22.9\times$ on the BERT-Base model compared to CPU implementations. In recent years, a number of cryptographic libraries supporting GPU accelerations have been released, e.g.,~\cite{ma2023secretflow,wang2023he,jawalkar2024orca,yang2024phantom}, showing great potential for improving runtime.
We believe future works should explore optimizing GPU-based frameworks for PTI to make real-time inference more feasible.

\textbf{Scalability on Generation Tasks.} Another concern is that existing PTI solutions are seldom evaluated on content generation tasks, e.g., translation and Q\&A. As shown in Table~\ref{tab:performance}, most studies have focused on classification tasks (e.g., GLUE and ImageNet) and achieved promising results. However, cryptographic solutions for transformers introduce approximation errors that vary by task. In classification tasks, evaluation primarily depends on identifying the highest category scores, where the errors may have minimal effect on overall accuracy~\cite{rovida2024transformer}.
Generation tasks could be more sensitive to such errors, as they can lead to nonsensical outputs and severely degrade performance. This issue was also highlighted in the open review\footnote{\url{https://openreview.net/forum?id=x3LxHdZX0f}} of PUMA~\cite{dong2023puma}. 
An example of a failed Q\&A is as follows:
\begin{itemize}
    \item Q: What is the largest animal?
    \item A: * a. What is the difference between? Sedition between? Sedition ? ? between
\end{itemize}

Thereafter, expanding PTI evaluations to include diverse generative applications (e.g., dialogue systems and creative writing tasks) will be essential for realizing their potential.

\textbf{Practicality in Real-World Applications.}
Despite their theoretical promise, the practicality of PTI solutions in real-world applications remains a significant challenge. Current implementations often involve high computational and memory overheads, making deployment on edge devices or in latency-sensitive scenarios impractical. Additionally, many cryptographic methods require extensive parameter tuning and complex setup, further hindering their usability. Future research should prioritize developing lightweight and user-friendly PTI frameworks that balance security and performance. Moreover, integrating these solutions into widely used AI platforms could facilitate their adoption in industries such as healthcare, finance, and personalized AI services, where privacy is paramount. Addressing these concerns will be crucial for bridging the gap between theoretical advancements and practical deployments.

\section*{Ethical Statement}

Our experiments use only open-source datasets and do not involve human participants or sensitive data. While our work promotes secure AI deployment, we acknowledge potential misuse and advocate for responsible AI practices.

\appendix
\bibliographystyle{named}
\bibliography{ijcai25}

\end{document}